\documentstyle[12pt]{article}

\begin{document}
\begin{titlepage}
\begin{center}

October 23, 2000     \hfill    LBNL-46870 \\

\vskip .5in

{\large \bf Reply to ``On Stapp's `Nonlocal Character of Quantum Theory' ''}
\footnote{This work is supported in part by the Director, Office of Science, 
Office of High Energy and Nuclear Physics, Division of High Energy Physics, 
of the U.S. Department of Energy under Contract DE-AC03-76SF00098}

\vskip .50in
Henry P. Stapp\\
{\em Lawrence Berkeley National Laboratory\\
      University of California\\
    Berkeley, California 94720}
\end{center}

\vskip .5in

\begin{abstract}

The question raised by Shimony and Stein is examined and used 
to explain in more detail a key point of my proof that any 
theory that conforms to certain general ideas of orthodox 
relativistic quantum field theory must permit transfers of 
information over spacelike intervals. It is also explained 
why this result is not a problem for relativistic quantum 
theory, but, on the contrary, opens the door to a satisfactory
realistic relativistic quantum theory based on the ideas 
of Tomonaga, Schwinger, and von Neumann.

\end{abstract}

\end{titlepage}
 
\newpage
\renewcommand{\thepage}{\arabic{page}}
\setcounter{page}{1}

Shimony and Stein[1] have raised a question about an essential claim made in 
my 1997 paper[2]. I begin by explaining the claim, and the question they
raised.

Lines 1 through 5 of my proof[2] show that under certain 
explicitly stated conditions the statement 
$$
L2\Rightarrow [(R2\wedge R2+)\rightarrow(R1\Box\rightarrow R1-)] \eqno (1)
$$
is true, while lines 6 through 14 of that proof show that 
under these same conditions the  statement 
$$
L1\Rightarrow [(R2\wedge R2+)\rightarrow(R1\Box\rightarrow R1-)] \eqno (2)
$$
is false. Shimony and Stein arrive at the same conclusion---namely 
that (1) is true and (2) is false---under similar 
conditions. I then claim that this fact, that (1) is true and (2) is 
false, entails that information must sometimes be transferred over 
space-like intervals. Shimony and Stein question this claim, and suggest 
that one must make a hidden-variable assumption, as was done in 
Bell's theorems [3,4], in order to arrive at this strong conclusion. 

This issue is important, because all the assumptions used my proof are 
elements of orthodox quantum philosophy, and hence my claim, if valid,
means that the precepts of orthodox quantum philosophy entail that 
information must sometimes be transferred over spacelike intervals. That 
conclusion is far stronger than what is proved by Bell's theorem[3],
and its usual generalizations[4,5], and it seems to have profound 
implications for development of relativistic quantum theory.

To provide an adequate foundation for the discussion I need to 
explain the meanings of (1) and (2), the assumptions that go into my 
proof that (1) is true and (2) is false, and the technical differences 
between my assumptions and those of Shimony and Stein.

The conditions under which I prove that (1) is true and
(2) is false are these:\\

A. The choices made by the experimenters in each of the two 
regions {\bf R} and {\bf L} about which experiment will be 
performed in that region can be treated as {\it free choices}, 
or {\it free variables}.

B. There is at least one Lorentz frame of reference, call it LF, 
such that if in that frame every point of the spacetime region {\bf L} 
is earlier than every point in the spacetime region {\bf R} then for any
experiment freely chosen and performed in the earlier region {\bf L} 
the outcome that appears to observers in that region can be taken to be
independent of which experiment will be freely chosen and performed 
later in the region {\bf R}: the universe can be regarded as evolving 
forward in time in LF and, in particular, there is no action of a 
free choice made {\it later} in {\bf R} upon an outcome that has 
{\it already appeared earlier} in {\bf L}. 
This assumption is called LOC1.

C. No matter which experiments are freely chosen and performed,
the predictions of quantum theory will be satisfied.

These assumptions are, I believe, compatible with the precepts of
orthodox quantum thinking, and are, in a broad sense, entailed
by them.

Notice that the truth of certain very special contrary-to-fact assertions 
is entailed by these assumptions. In particular, if the set of
possible worlds is limited by conditions A, B, and C then,\\

SF: For any possible world $W$, the following statement is true:\\
If the situation in $W$ is such that\\
1. The Hardy experimental conditions are satisfied,\\ 
2. Experiment L2 is freely chosen and performed in {\bf L},\\ 
3. Experiment R2 is freely chosen and performed in {\bf R}, and\\ 
4. The outcome $L2+$ appears in {\bf L},\\ 
then in any possible world $W'$ that is the same as world $W$ 
except for possible consequences choosing and performing in 
region {\bf R} the experiment R1, instead of the experiment 
chosen and performed in {\bf R} in world $W$, the outcome 
in the earlier region {\bf L} is $L2+$.

The result asserted by SF is immediately entailed by the stated 
assumptions, actually just A and B, and it is expressed symbolically as  
$$
(L2 \wedge R2 \wedge L2+)\Rightarrow (R1\Box \rightarrow L2+).\eqno(3)
$$

This statement asserts, in brief, that if the theoretical 
conditions A, B, and C are satified then freely choosing 
and performing R1, instead of R2, in the later region {\bf R},
leaves the earlier outcome in {\bf L} undisturbed. 

Similarly, the symbolic statement (1) asserts that, under the 
three conditions A, B, and C, if experiment L2 is freely chosen 
and performed in the earlier region {\bf L} then a certain statement 
SR is true, whereas statement (2) makes the same claim under
condition L1. Thus the conjunction of the facts that (1) is true 
and (2) is false implies that the truth of statement SR depends 
nontrivially on which of the two alternative possible experiments, 
L1 or L2, is freely chosen and performed in the space-time 
region {\bf L}.

The statement SR just mentioned is represented symbolically by
$$
[(R2\wedge R2+)\rightarrow(R1\Box\rightarrow R1-)] \eqno (4)
$$
It is an assertion about a possible world $W$, and it states \\

\noindent SR: If in the possible world $W$ the experimenter in 
the space-time region {\bf R} freely chooses and 
performs experiment R2 and gets the outcome $R2+$, then 
in any possible world $W'$ that is the same as world $W$
except for the possible consequences of choosing in the 
region {\bf R} the experiment R1, instead of
whatever was chosen in $W$ (namely R2), the outcome in
{\bf R} is $R1-$.

In reference [2] I justified each step in the proof that 
statement (1) is true and  statement (2) is false by using 
the machinery of David Lewis's rules of reasoning with 
counterfactual statements. The Lewis machinery is reasonable 
and orthodox, but was created in the climate where the ideas 
of deterministic classical physics prevailed, and in the end 
it is merely a set of conventions designed to cope in a 
deterministic setting with the idea that something other than
what actually happens 'could have happened'. The conventions
are designed to mesh with our intuitions about the 
proper use of contrary-to-fact statements, but there are
other contending rules, and the whole situation is somewhat
controversial. But as I have emphasized, and Shimony and 
Stein have agreed, the quantum situation permits a more 
direct approach, which avoids leaning on the basically 
conventional features of the classical approach. Instead, 
one can exploit the fact the concept of a `free choice' is 
compatible with quantum theory, due to its basically 
indeterministic character. This allows one to stay with 
ordinary logic plus a natural specified meaning for the needed 
counterfactual assertion. 

In order to have a common ground for dealing with the 
concerns of Shimony and Stein I shall, in this paper, adopt 
this alternative approach, which is strictly line with 
quantum thinking, rather than relying on Lewis's classical 
rules. However, apart from  this technical change, I shall 
adhere to the logical form of the argument that I used in 
reference [2]. In particular, I shall retain the following
natural meaning of the statement 
$(E\Box \rightarrow O)$:\\

$(E\Box \rightarrow O)$ is, by definition, true in a possible world 
$W$ if and only if outcome $O$ occurs in any possible world 
$W'$ that is the same as world $W$ except for the possible 
consequences of freely choosing and performing experiment E 
instead of the alternative experiment freely chosen and performed 
in $W$. The set of possible worlds is limited by the 
specified conditions A, B, and C.

To use this definition one must limit ``the possible 
consequences of ...'' This is always done by using LOC1. 
Since this definition is toothless without this condition LOC1, 
or some such condition, and since LOC1 is used only in 
connection with this definiion, it is not unreasonable to 
incorporate LOC1 into the definition of the counterfactual 
statement. Shimony and Stein have done essentially that. 
However, they did not do {\it exactly} that. My condition
LOC1 excludes from the effects of changing a free 
choice {\it only} effects on outcomes that have already
appeared {\it earlier}, in the special Lorentz frame LF. 
But Shimony and Stein exclude all effects that lie outside the 
forward light-cone of the region in which the change in 
the free choice occurs. 

I use the weaker assumption LOC1 because the truth of LOC1 is certainly
compatible with the principles of relativistic quantum field 
theory, and is in fact entailed by them, whereas the stronger form used
by Shimony and Stein is incompatible those principles. It is much 
clearer to argue directly from assumptions that are true, in the sense 
of being consequences of orthodox quantum theory, rather that making an 
assumption that is incompatible with relativistic quantum theory.

The fact that LOC1 is entailed in orthodox relativistic quantum
field theory is proved by noting that the possibility of defining
{\it one} such frame  LF follows from the Tomonaga-Schwinger [6,7] formulation 
of relativistic quantum field theory, in which advancing space-like 
surfaces are the analogs of the advancing constant-time surfaces 
of the non-relativistic formulation of von Neumann [8]. Of course,
an infinitude of {\it alternative} possible choices for LF
can be found: {\it any} frame will do. But the required property 
follows for {\it only one frame {\bf or} another}, not for any two 
or more together.

With the stage thus set, I can turn to the central question 
of whether the conjunction of the truth of (1) and the 
falsity of (2) can be reconciled, as Shimony and Stein appear to 
suggest, with the idea that {\it no} information about the choice made 
in region {\bf L} can get to region {\bf R}, which is situated 
spacelike relative to {\bf L}. 

To see the apparent conflict one can consider the consequence
of the fact that (1) is true and (2) is false in the context 
of the orthodox idea that ``nature chooses the outcome'' of 
the experiment chosen by the experimenter. In this context 
the consequence of the truth of (1) and falsity of (2) is that
SR asserts the existence of a definite theoretical connection 
between the outcomes that nature delivers under the two alternative
possible conditions, and that this theoretically necessary 
condition on what nature can do in {\bf R} depends 
nontrivially on which experiment is freely chosen and 
performed by the experimenter in $L$. 

But how can {\it any} theoretical model---hidden-variable or not---
fulfill conditions on Nature's choices in region {\bf R}
that depend nontrivially on which free choice is 
made in {\bf L} if no information about this choice made 
in {\bf L} can be present in {\bf R}?

This  apparent result, that any theoretical model
that conforms to the conditions A, B, and C 
must accomodate transfers of information over a 
spacelike intervals, does not conflict with the 
requirements of the theory of relativity, {\it in 
the context of quantum theory}: it conflicts only 
with a certain prejudice generated by uncritically 
extending to indeterministic quantum theory a feature of 
its deterministic classical approximation. This prejudice 
has, in fact, been the barrier that has blocked for many years 
the creation of a satisfactory realistic formulation of 
relativistic quantum theory. In a quantum context the 
Lorentz requirements of relativity theory pertain exclusively
to relationships among observables, not to the reality that
lies behind the phenomena. Thus the obvious realistic, 
relativistic quantum theory is just relativistic 
quantum field theory with a preferred sequence of advancing 
Tomonaga-Schwinger [6,7] spacelike surfaces defining the 
successive instants ``now''. 

This entails, of course, a 
reversion to the pre-relativity Newtonian idea of an 
absolute time, or something similar to it, at the underlying 
ontological level. But the founders of quantum theory  
strongly stressed the fact that this theory, as they 
conceived it, was only about relationships between 
observations, not about properties of the underlying reality. 
The Tomonage-Schwinger theory maintains all the observable 
requirements of the theory of relativity, no matter how the 
preferred sequence of advancing spacelike surfaces 
is chosen. Hence the only thing actually blocking acceptance 
of this theory as the relativistic quantum theory of reality 
is the prejudicial assumption that the reality itself, like 
the connections between observations, can have no transfer of 
information over spacelike intervals. But the fact that 
this condition can be maintained in the {\it deterministic} 
classical limit, where the entire history of the universe 
is determined by the initial conditions, and can immediately 
be laid out on a space-time background, with no free choices 
allowed, does not entail that it can be maintained in the 
full indeterministic theory with free choices allowed. 

The analysis of the Hardy case supports the view that the 
reality behind the indeterministic quantum rules cannot 
maintain this constraint. That observation immediately 
elevates John von Neumann's [8] formulation of quantum theory,
applied to Tomonaga-Schwinger relativistic quantum field 
theory, to prime candidacy as the paradigm relativistic 
quantum theory of reality.\\

Shimony and Stein allege that this apparent result---that 
the information about whether L1 or L2 was freely
chosen and performed in region {\bf L} must be available in region
{\bf R} of---is incorrect. They base their argument on the assertion 
that {\it the semantical truth conditions for the counterfactual in 
question refer explicitly to the entire exterior of the extended future 
light-cone of {\bf R}}.

That claim about the {\it entire} exterior is not exactly true 
in my version of the proof. The statement SR combine with LOC1 says:\\

\noindent SR-LOC1: ``If in the possible world $W$ the experiment 
R2 is freely chosen and performed in {\bf R} and the outcome 
there is $R2+$ then if $W'$ is a possible world that is the 
same as $W$ in {\bf L}, but in which R1 is freely chosen and 
performed in {\bf R}, instead of R2, the outcome in {\bf R} 
in world $W'$ is $R1-$. 

In spite of the difference between the light-cone version
of the causality condition used by Shimony and Stein and
the condition LOC1 used by me, this combined statement SR-LOC1
exhibits the feature pointed to by Shimony and Stein: a reference 
to the region {\bf L}, which lies outside the forward light cone 
of the region {\bf R}. It is this implicit reference of SR to {\bf L} 
that Shimony and Stein are concerned about. The question is whether 
this reference to {\bf L} upsets my essential claim that the conjunction 
of the truth of (1) and the falsity of (2) requires the information 
about whether L1 or L2 is performed in {\bf L} to be present in 
{\bf R}.

Let me begin my answer by explaining the question in more detail.

The statement SR  involves the words ``instead
of''. We have a clear idea of what we mean here
by ''instead of''. In the real situation the experimenter
in {\bf R} makes the choice R2. But we have assumed that, just
at the moment of choosing, the other choice R1 could have popped
out instead of R2. But the central idea is that everything 
{\it prior to that moment of choosing} is exactly what it is 
in the actual world: there is just {\it one} evolving quantum world, 
which could go either way at the moment of choice. 

This condition of {\it sameness  prior to the moment of choice} 
is the condition that limits the changes permitted by the phrase 
``except for the possible consequences of the change in 
the free choice'': no possible consequence of a changed choice can lie 
earlier than the moment of choice.
 
The point raised by Shimony and Stein, as applied to my 
argument, is that this implicit reference to the (unchanged) 
state of affairs (in {\bf L}) prior to the moment of the choice 
between R1 and R2 is an essential element of the very idea 
of ``instead of'' that appears in the statement SR. Hence 
there is in SR an essential implicit reference to region {\bf L}, 
even though all the symbols explicitly appearing in SR pertain
to possible events in {\bf R}.

Their concern about this implicit reference to {\bf L} stems from 
the fact that in my 1997 paper I based my argument---for the claim that the 
conjunction of the truth of (1) and falsity of (2) entails a 
violation of the idea that ``observable effects can propagate 
only into the future (light-cone)''---on the fact that 
``everything mentioned in SR is an observable phenomenon in region 
{\bf R}.'' Their concern is that the essential implicit reference
of SR to the region {\bf L} might upset my argument.

This essential implicit reference of SR to {\bf L} does not
affect my argument. To understand why it does not, one must
note that the steps in a logical argument are like a series
of black boxes, each of which displays explicitly only certain
of the variables of the system. These explicitly displayed variables
are like inputs and outputs: certain connection between
these variables are exhibited, but the reasons why these connections
hold are not shown. However, all of the relevant effects pertaining to
the inner workings must be controlled by the displayed variables. 

In the statement (1),
$$
L2\Rightarrow [(R2\wedge R2+)\rightarrow (R1\Box\rightarrow R1-)],
$$
the only displayed  variables are $L2, R2, R2+, R1,$ and $R1-$.
The input conditions are  $L2, R2, R2+,$ and $R1$, and the output is $R1-$.
The statement asserts that if the input variables  $L2, R2, R2+,$ and $R1$ 
are put into a certain logical expression, the output must be $R1-$,
never $R1+$, But the falseness of (2) says that if the inputs are changed
only by changing L2 to L1, then the output is no longer restricted to $R1-$:
it is now allowed to be $R1+$. So changing the input variable from L2
to L1 has affected the output variable $R1+/R1-$. There can be all 
sorts of dependence on all sorts of inner variables, but whatever these
dependences are they {\it must}, to the extent that they are relevant
to the output conclusion,  be controlled by the input variables, if 
the statement is indeed logically correct. So, in this case at hand, 
changing the input variable L1/L2 affects nontrivially the output 
variable $R1+/R1-$. But then the information about whether L1 or L2 is 
chosen in {\bf L} must get to the region {\bf R} where the value of the 
output variable $R1+/R1-$ is displayed.\\

\noindent {\bf References}

1. Abner Shimony and Howard Stein, American Journal of Physics.

2. Henry P. Stapp, American Journal of Physics, {\bf 65}, 300-304 (1997).

3. John Bell, Physics {\bf 1}, 195 (1964).

4. John Bell, Proc. Int. School of Pysics `Enrico Fermi', course II,\\
   New York, Academic, 171 (1971). 

5. John F. Clauser and Abner Shimony, Rep. Prog. Phys. {\bf 41}, 1881 (1978)

6. Sin-itiro Tomonaga, Progress of Theoretical Physics, {\bf 1}, 27 (1946).
 
7. Julian Schwinger, Physical Review, {\bf 82}, 914 (1951).
  
8. John von Neumann, {\it Mathematical Foundations of Quantum Theory,}\\
   Princeton Univ. Press, Princeton NJ, 1955.
  
\end{document}